\documentclass[journal]{IEEEtran}

\ifCLASSINFOpdf
\else
   \usepackage[dvips]{graphicx}
\fi
\usepackage{url}

\usepackage{tikz}
\usepackage{amsmath,amsfonts,amssymb}
\usepackage{bm}
\usepackage{graphicx,graphics}
\usepackage{cases}
\usepackage[noadjust]{cite}
\usepackage{color}
\usepackage{cite,url}
\usepackage{verbatim}
\usepackage{balance}
\usepackage{multirow}
\usepackage{stfloats}
\usepackage{xspace}
\usepackage{amsthm}
\usepackage{mathtools}
\usepackage{subfig}
\usepackage{caption}
\usepackage{algorithm}
\usepackage{algpseudocode}
\usepackage{float}
\usepackage{enumerate}
\usepackage{enumitem}
\floatname{algorithm}{Procedure}

\graphicspath{{Figures/}}

\newcommand{\innerpS}[2]{\left \langle {#1} , {#2} \right \rangle_{\untsph}}
\newcommand{\innerpSO}[2]{\left \langle {#1} , {#2} \right \rangle_{\SO}}

\newcommand{\lsphl}{\mathcal{H}_{L}}
\newcommand{\untsph}{\mathbb{S}^{2}}
\newcommand{\lsph}{L^{2}(\untsph)}
\newcommand{\SO}{\mathbb{SO}(3)}
\newcommand{\lSO}{L^{2}(\SO)}
\newcommand{\bv}[1]{\boldsymbol{#1}}
\newcommand{\unit}[1]{\bv{\hat{#1}}}
\newcommand{\intsph}{\int_{\mathbb{S}^{2}}}
\newcommand{\dfn}{\triangleq}
\newcommand{\D}{\mathcal{D}}

\newcommand{\B}[1]{\mathbf{#1}}

\newcommand{\E}{\mathbb{E}}
\newcommand{\mse}{\mathcal{E}_{ms}}
\newcommand{\convSO}{\circledast}
\newcommand{\const}{c_{\ell}}

\newcommand{\figref}[1]{Fig.\,\ref{#1}}
\newcommand{\secref}[1]{Section\,\ref{#1}}

\newcommand{\R}{\mathbb{R}}

\newcommand{\res}{j}

\newcommand{\wav}[1]{\Psi^{(#1)}}
\newcommand{\scal}{\Phi}
\newcommand{\wavcoef}[2]{w_{#1}^{\wav{#2}}}
\newcommand{\scalcoef}[1]{w_{#1}^\scal}

\newcommand{\tilwav}[1]{\Gamma_{\Psi}^{(#1)}}
\newcommand{\tilscal}{\Gamma_{\Phi}}
\newcommand{\dir}[2]{\zeta_{#1}^{#2}}
\newcommand{\dil}{\gamma}

\newcommand{\snr}[1]{\mathrm{SNR}^{#1}}

\DeclarePairedDelimiterX\abs[1]{\lvert}{\rvert}{#1}
\DeclarePairedDelimiterX\parn[1]{(}{)}{#1}
\DeclarePairedDelimiterX\set[1]{\lbrace}{\rbrace}{#1}
\DeclarePairedDelimiterX\norm[1]{\lVert}{\rVert}{#1}
\DeclarePairedDelimiterX\brak[1]{\lbrace}{\rbrace}{#1}
\DeclarePairedDelimiterX\coeff[1]{(}{)}{#1}

\newcommand{\displayskipshrink}{%
	\setlength{\abovedisplayskip}{3.0pt plus 1.0pt minus 1.0pt}
	\setlength{\abovedisplayshortskip}{0pt plus 1.0pt minus 1.0pt}
	\setlength{\belowdisplayskip}{3.0pt plus 1.0pt minus 1.0pt}
	\setlength{\belowdisplayshortskip}{3.0pt plus 1.0pt minus 1.0pt}}

\newtheorem{remark}{Remark}
\newtheorem{theorem}{Theorem}

\newcommand{\ad}[1]{\textcolor{blue}{#1}}

\title{Multiscale Optimal Filtering on the Sphere}

\author
{
	Adeem Aslam, \IEEEmembership{Student Member, IEEE}, Zubair Khalid, \IEEEmembership{Senior Member, IEEE} and Jason D. McEwen, \IEEEmembership{Senior Member, IEEE}
	\thanks
	{
		A.~Aslam and Z.~Khalid are with the School of Science and Engineering, Lahore University of Management Science, Lahore, Pakistan (e-mail: adeem.aslam@lums.edu.pk, zubair.khalid@lums.edu.pk).
	}
	\thanks
	{
		J.~D.~McEwen is with the Mullard Space Science Laboratory, University College London, Surrey, England (e-mail: jason.mcewen@ucl.ac.uk).
	}
}

\begin{document}

\displayskipshrink

\maketitle

\begin{abstract}
We present a framework for the optimal filtering of spherical signals contaminated by realizations of an additive, zero-mean, uncorrelated and anisotropic noise process on the sphere. Filtering is performed in the wavelet domain given by the scale-discretized wavelet transform on the sphere. The proposed filter is optimal in the sense that it minimizes the mean square error between the filtered wavelet representation and wavelet representation of the noise-free signal. We also present a simplified formulation of the filter for the case when azimuthally symmetric wavelet functions are used. We demonstrate the use of the proposed optimal filter for denoising of an Earth topography map in the presence of additive, zero-mean, uncorrelated and white Gaussian noise, and show that the proposed filter performs better than the hard thresholding method and weighted spherical harmonic~(weighted-SPHARM) signal estimation framework.
\end{abstract}


\section{Introduction}
\label{sec:intro}
Signals exhibiting angular dependence are naturally defined on the surface of the sphere and are called spherical signals. Such signals arise in different fields of science and engineering, such as astronomy~\cite{Jarosik:2011,Galanti:2019}, cosmology~\cite{Dahlen:2008,Marinucci:2008}, geodesy~\cite{Simons:2006Polar}, computer graphics~\cite{Lam:2006,nadeem2016spherical}, medical imaging~\cite{Chung:2007,Bates:2016}, acoustics~\cite{Zhang:2012,Bates:2015} and wireless communication~\cite{Alem:2015,bashar:2016}. Signal acquisition in all of these application areas is influenced by unwanted, yet unavoidable, noise which places signal estimation from its noise-contaminated samples at the heart of signal processing techniques. There is an abundance of literature on signal filtering and estimation methods~\cite{hlawatsch:2000,sasgen2006wiener,Yeo:2008,McEwen:2008,klees2008design,arora2010optimal,McEwen:2012,Khalid:2013,sadeghi2014}. Some of these methods process signals using spatial~(temporal) or spectral representation while others use joint spatial~(temporal)-spectral representation. 

Signal filtering and estimation has also been carried out using the joint scale-space representation in the Euclidean domain~\cite{Starck:1994,Renaud:2005}. Such a representation is enabled by the wavelet transform~\cite{Mallat:1989,Daubechies:1990,Mallat-book:2009}, which has proved invaluable in the multiscale analysis of one dimensional and multidimensional Euclidean domain signals. The framework for wavelet transform on the sphere has been extensively investigated in the literature as well~\cite{Narcowich:1996,Freeden:1997,Antoine:1998:wavelets,Antoine:1999,Wiaux:2005,Starck:2006,Wiaux:2008,McEwen:2018}. Since dilation on the sphere, unlike in the Euclidean setting, can be defined in more than one way, there are different formulations and algorithms for multiscale analysis of spherical signals~\cite{Antoine:1999,Wiaux:2005,Wiaux:2008,McEwen:2018}.

Signal estimation using wavelet transform is based on the observation that noise has a distributed representation in the wavelet domain whereas signals of interest are typically sparsely represented, which can be exploited using different thresholding methods~\cite{Starck:1999,Leistedt:s2let_axisym}. Multiscale signal estimation in the Euclidean domain has also been carried out using signal and noise statistics through Wiener filtering~\cite{Starck:1994}. In this work, we adopt the philosophy of Wiener filtering to propose a multiscale filter for signal estimation by using the scaled-discretized wavelet transform on the sphere~\cite{Wiaux:2008,McEwen:2018}. The designed filter is optimal in the sense that the filtered wavelet representation is the minimum mean square error estimate of the scale-discretized wavelet transform of the noise-free signal. Before presenting the design of the optimal filter, we review the fundamentals of signal analysis on the sphere and $\SO$ rotation group in \secref{sec:maths}. We briefly review the scale-discretized wavelet transform and present the optimal filtering framework in \secref{sec:filter_design}. In \secref{sec:analysis}, we demonstrate the performance of the proposed multiscale optimal filter by estimating a bandlimited Earth topography map in the presence of additive, zero-mean, uncorrelated and white Gaussian noise. We also test the robustness of the proposed multiscale filter, at different noise levels, against the hard thresholding method and the weighted spherical harmonic~(weighted-SPHARM) signal estimation framework~\cite{Chung:2007}, before concluding in \secref{sec:conclusion}.

\section{Mathematical Preliminaries}
\label{sec:maths}

\subsection{Signals on $2$-Sphere}
Spherical domain, also referred to as $2$-sphere~(or just sphere) is defined as $\untsph \dfn \{\unit{x} \in \R^3 : |\unit{x}| = 1\}$, where $|\cdot|$ represents the Euclidean norm. A point $\unit{x}\in\untsph$ can be represented as $\unit{x}(\theta,\phi) \dfn (\sin\theta\cos\phi, \sin\theta\sin\phi, \cos\theta)^{\mathrm{T}}$, where $(\cdot)^{\mathrm{T}}$ denotes vector transpose and the angles $\theta \in [0,\pi]$ and $\phi \in [0,2\pi)$ are referred to as colatitude and longitude, and are measured from the positive $z$-axis and positive $x$-axis~(in the $x-y$ plane) respectively. Complex-valued and square-integrable functions defined on the sphere form a Hilbert space $\lsph$, which is equipped with the following inner product for $f,g\in\lsph$
\begin{equation}
	\innerpS{f}{g} \dfn \int_{\untsph} f(\unit{x}) \overline{g(\unit{x})} ds(\unit{x}), \quad \int_{\untsph} \dfn \int_{\theta=0}^{\pi} \int_{\phi=0}^{2\pi},
	\label{eq:inner_product_S2}
\end{equation}
where $\overline{( \cdot )}$ denotes complex conjugate and $ds(\unit{x}) = \sin\theta d\theta d\phi$. Norm of the function $f$ is induced as $\norm{f}_{\untsph} = \sqrt{\innerpS{f}{f}}$ and its energy\footnote{We refer to finite energy functions as signals.} is given by $\innerpS{f}{f}$. For $\lsph$, spherical harmonics, denoted by $Y_{\ell}^m(\unit{x})$ for integer degree $\ell \geq 0$ and integer order $|m| \leq \ell$, form a complete set of orthonormal basis functions~\cite{Kennedy-book:2013}, which can be used to represent any signal $f \in \lsph$ as
\begin{equation}
	f(\unit{x}) = \sum\limits_{\ell,m}^{\infty}  (f)_{\ell}^m Y_{\ell}^m(\unit{x}), \quad \sum\limits_{\ell,m}^{\infty}\equiv\sum\limits_{\ell = 0}^{\infty} \sum\limits_{m = -\ell}^{\ell},
	\label{eq:Fourier_expansion_S2}
\end{equation}
where $(f)_{\ell}^m \dfn \innerpS{f}{Y_{\ell}^m}$ is the spherical harmonic~(spectral) coefficient of degree $\ell$ and order $m$, and constitutes the spectral domain representation of the signal $f$. Signal $f$ is considered bandlimited to degree $L$ if $(f)_{\ell}^m = 0$ for $\ell \geq L, |m| \le \ell$. Set of all such bandlimited signals forms an $L^2$-dimensional subspace of $\lsph$, denoted by $\lsphl$.

\subsection{Signals on $\SO$ Rotation Group}
Rotations on the sphere are specified by three Euler angles namely, $\omega \in [0,2\pi)$, $\vartheta \in [0,\pi]$ and $\varphi \in [0,2\pi)$, using the right handed $zyz$ convention, in which a point on the sphere is sequentially rotated by $\omega$, $\vartheta$ and $\varphi$ around $z$, $y$ and $z$ axes respectively. Group of all proper rotations\footnote{An improper rotation is a reflection about either an axis or the center of the spherical coordinate system.}, specified by the three Euler angles $(\varphi,\vartheta,\omega)$, is called the Special Orthogonal group, denoted by $\SO$. Each element in the $\SO$ rotation group is parameterized by the $3$-tuple of Euler angles $\rho = (\varphi,\vartheta,\omega)$. We define rotation of signals on the sphere through an operator $\D_{\rho}$. Spectral representation of a signal $f \in \lsph$ under the action of $\D_{\rho}$ is given by
\begin{align}
	(\D_{\rho} f)_{\ell}^m \dfn \innerpS{\D_{\rho} f}{Y_{\ell}^m} = \sum_{m'=-\ell}^{\ell} D^{\ell}_{m,m'}(\rho) (f)_{\ell}^{m'},
	\label{eq:rotated_flms}
\end{align}
where $D^{\ell}_{m,m'}(\rho)$ is called the Wigner-$D$ function of integer degree $\ell \ge 0$ and integer orders $|m|,|m'| \le \ell$~\cite{Kennedy-book:2013}.

For square-integrable and complex-valued functions, of the form $g(\rho), \nu(\rho)$, defined on the $\SO$ rotation group, the inner product is given by
\begin{align}
	\innerpSO{g}{\nu} \dfn \int_{\SO} \!\!\! g(\rho) \overline{\nu(\rho)} \, d\rho, \quad \int_{\SO} \!\!\!\!\! \dfn \int\limits_{\varphi=0}^{2\pi} \int\limits_{\vartheta=0}^{\pi} \int\limits_{\omega=0}^{2\pi},
	\label{eq:inner_product_SO3}
\end{align}
where $d\rho = d\varphi \sin\vartheta d\vartheta d\omega$. Equipped with the inner product in \eqref{eq:inner_product_SO3}, the set of signals defined on the $\SO$ rotation group form a Hilbert space, denoted by $\lSO$, which has Wigner-$D$ functions as basis functions that satisfy the following orthogonality relationship~\cite{Kennedy-book:2013}
\begin{align}
	\innerpSO{D^{\ell}_{m,m'}}{D^{p}_{q,q'}} \!\!\! = \const \, \delta_{\ell,p} \delta_{m,q} \delta_{m',q'}, \, \const \dfn \frac{8\pi^2}{2\ell+1}.
	\label{eq:WignerD_ortho}
\end{align}
Hence, we can express any signal $g \in \lSO$ as
\begin{equation}
	g(\rho) = \!\!\!\sum_{\ell,m,m'}^{\infty} (g)^{\ell}_{m,m'} \overline{D^{\ell}_{m,m'}(\rho)}, \quad \sum\limits_{\ell,m,m'}^{\infty} \!\!\! \equiv \sum\limits_{\ell=0}^{\infty} \sum\limits_{m=-\ell}^{\ell} \sum\limits_{m'=-\ell}^{\ell},
	\label{eq:Fourier_expansion_SO3}
\end{equation}
where $(g)^{\ell}_{m,m'} = \left(1/\const\right) \innerpSO{g}{\overline{D^{\ell}_{m,m'}}}$ forms the spectral domain representation of the signal $g$.

\section{Optimal Filtering using Scale-Discretized Wavelet Transform}
\label{sec:filter_design}
We consider a realization of an anisotropic random process as the source signal $s(\unit{x})$, which is contaminated by a realization of an additive, zero-mean and anisotropic random process, called the noise signal $z(\unit{x})$. The source and noise signals are assumed to be uncorrelated, i.e., $\E\{(s)_{\ell}^m \overline{(z)_{p}^q}\} = 0, \, \forall \, \ell, p, |m| \le \ell, |q| \le p$, where $\E\{\cdot\}$ represents the expectation operator. We further assume that the spectral covariances of the source and noise processes, denoted by $\B{C}^s$, $\B{C}^z$ respectively and defined as $C^s_{\ell m,p q} = \E\{(s)_{\ell}^m \overline{(s)_{p}^q}\}$, $C^z_{\ell m,p q} = \E\{(z)_{\ell}^m \overline{(z)_{p}^q}\}$, are known. Given the noise-contaminated observation $f(\unit{x}) = s(\unit{x}) + z(\unit{x})$, the problem under consideration is to obtain an estimate of the source signal, denoted by $\tilde{s}(\unit{x})$, by filtering the observation $f$ at different scales using the scale-discretized wavelet transform. Before we design an optimal filter for signal estimation, we present a multiscale representation of signals in the wavelet domain.

\subsection{Scale-Discretized Wavelet Transform on $2$-Sphere}
\label{sec:wavelets_RKHS}
In this section, we briefly review the directional scale-discretized wavelet transform given in~\cite{Wiaux:2008,McEwen:2018}. For a signal $f \in \lsph$, the directional scale-discretized wavelet transform is defined as
\begin{align}
	\wavcoef{f}{\res}\!(\rho) \! &\dfn \! \innerpS{f}{\D_{\rho}\wav{\res}} \!\!\! = \!\!\!\! \sum_{\ell,m,m'}^{\infty} \!\!\! (f)_{\ell}^m \overline{(\wav{\res})_{\ell}^{m'}} \, \overline{D^{\ell}_{m,m'}(\rho)},
	\label{eq:wavelet_coeff}
\end{align}
where $\wavcoef{f}{\res} \in \lSO$ is the wavelet coefficient of the signal $f$, $\res$ is the discrete wavelet scale, $(\wav{\res})_{\ell}^{m'}$ is the spectral coefficient of the wavelet function $\wav{\res} \in \lsph$ and we have used \eqref{eq:rotated_flms} along with the orthonormality of spherical harmonics to evaluate the second equality. Wavelet functions capture the high frequency~(rapid variations along the space) content of the signal. Low frequency content of the signal is represented by the scaling coefficient\ad{,} defined as
\begin{align}
	\scalcoef{f}(\unit{x}) = \innerpS{f}{\D_{\unit{x}}\scal} = \!\! \sum_{\ell,m}^{\infty} \sqrt{\frac{\const}{2\pi}} (f)_{\ell}^m \overline{(\scal)_{\ell}^0} \,\, Y_{\ell}^m(\unit{x}),
	\label{eq:scaling_coeff}
\end{align}
for the azimuthally symmetric scaling function $\scal \in \lsph$\footnote{For an azimuthally symmetric function $f$, $(f)_{\ell}^m = 0, \forall m \ne 0$.}, where $\D_{\unit{x}} \dfn \D(\varphi,\vartheta,0)$ and we have used the following relation between Wigner-$D$ functions and spherical harmonics~\cite{Kennedy-book:2013}
\begin{align}
	D^{\ell}_{m,0}(\varphi,\vartheta,0) = \sqrt{\frac{4\pi}{2\ell+1}} \overline{Y_{\ell}^m(\vartheta, \varphi)}.
	\label{eq:Wigner_SphHarm_relation}
\end{align}
Signal $f$ can be reconstructed perfectly from its wavelet and scaling coefficients using the following expression
\begin{align}
	f(\unit{x}) \!&= \!\!\! \displaystyle\int\limits_{\untsph} \!\!\! \scalcoef{f}\!(\unit{y}) (\D_{\unit{y}}\scal)\!(\unit{x}) ds(\unit{y}) \! + \!\!\!\!\!\! \displaystyle\int\limits_{\SO} \!\! \sum_{\res=\res_1}^{\res_2} \!\! \wavcoef{f}{\res}\!\!(\rho) (\D_{\rho}\!\wav{\res})\!(\unit{x}) d\rho,
	\label{eq:f_recons_wavelet}
\end{align}
provided the following admissibility condition holds
\begin{align}
	\const \left( \frac{1}{2\pi} \left|(\scal)_{\ell}^0\right|^2 \!\!\! + \sum_{\res=\res_1}^{\res_2} \sum_{m'=-\ell}^{\ell} \left|\big(\wav{\res}\big)_{\ell}^{m'}\right|^2 \right) \!\! = 1, \,\, \forall \,\, \ell,
	\label{eq:admis_cond}
\end{align}
where $\res_1$ and $\res_2$ denote the minimum and maximum wavelet scale\footnote{Wavelet scale is inversely proportional to the degree $\ell$.} respectively and this condition can be derived from \eqref{eq:f_recons_wavelet} using \eqref{eq:wavelet_coeff} and \eqref{eq:scaling_coeff}. Wavelet and scaling functions, which satisfy the admissibility condition, have the following spectral representation
\begin{align}
	\begin{split}
		\left(\wav{\res}\right)_{\ell}^{m} \!\!=\!\! \frac{1}{\sqrt{\const}} \, \tilwav{\res}(\ell,\dil) \, \dir{\ell}{m}, \,
		(\scal)_{\ell}^{0} \!=\! \sqrt{\frac{2\pi}{\const}} \, \tilscal(\ell,\dil),
	\end{split}
	\label{eq:wavelet_scaling_spectral_functions}
\end{align}
such that $\sum_{m=-\ell}^{\ell} |\dir{\ell}{m}|^2 = 1$ for all values of $\ell$ for which $\dir{\ell}{m}$ is non-zero for at least one value of $m$, and	$\left|\tilscal\right|^2 + \sum_{\res=\res_1}^{\res_2} \left|\tilwav{\res}\right|^2 = 1$. Here $\dil$ is the harmonic space dilation parameter, $\tilwav{\res}$, $\tilscal$ are the harmonic tiling functions which control the angular localization, and $\dir{\ell}{m}$ encodes the directional features of the wavelet functions.\footnote{See \cite{Wiaux:2008,McEwen:2018} for details on the construction of harmonic tiling functions and directionality component of the wavelet functions.}

\vspace{-2.5mm}
\subsection{Optimal Filtering in the Wavelet Domain}
We define a joint $\SO$-scale domain filter function as
\begin{align}
	\Xi(\rho;\res) = \sum_{\ell,m,m'}^{L_{\Xi_{\res}}-1} \left(\Xi(\cdot;\res)\right)^{\ell}_{m,m'} \overline{D^{\ell}_{m,m'}(\rho)},
	\label{eq:Xi}
\end{align}
for wavelet scales $\res = \res_1,\ldots,\res_2$. Action of this filter on wavelet coefficients of the noise-contaminated observation $f$ is given by $\SO$ convolution, defined as~\cite{Kostelec:2008,Vilenkin-book:1968}
\begin{align*}
	\wavcoef{\tilde{s}}{\res}\!(\rho) \! = \! \left(\Xi(\cdot;\res) \convSO \wavcoef{f}{\res}\right)\!(\rho) = \!\!\!\!\! \displaystyle \int\limits_{\SO} \!\!\! \Xi(\rho \rho_1^{-1};\res) \wavcoef{f}{\res}\!(\rho_1) d\rho_1,
\end{align*}
where $\tilde{s}(\unit{x})$ is the source signal estimate and $\convSO$ denotes convolution on the $\SO$ rotation group. Using the addition formula for Wigner-$D$ functions~\cite{Marinucci-book:2011}, $\SO$ convolution can be written in a computationally more amenable form as~\cite{Vilenkin-book:1968}
\begin{align}
	\wavcoef{\tilde{s}}{\res}(\rho) &= \!\! \sum_{\ell,m,m'}^{L-1} \!\! \const \! \sum_{k=-\ell}^{\ell} \! (f)_{\ell}^k \left(\Xi(\cdot;\res)\right)^{\ell}_{m,k} \overline{(\wav{\res})_{\ell}^{m'}} \overline{D^{\ell}_{m,m'}(\rho)},
	\label{eq:filtered_wavelet_coeff}
\end{align}
where $\const$ is given in \eqref{eq:WignerD_ortho}, $L$ is the bandlimit of signal $f$ and we have assumed, without loss of generality, that bandlimit of the filter function at each scale is equal to the signal bandlimit, i.e., $L_{\Xi_\res} = L$. We design an optimal filter that minimizes the joint $\SO$-scale domain mean square error given by
\begin{align}
	\mse &= \E\left\{\sum_{\res = \res_1}^{\res_2} \norm{\wavcoef{\tilde{s}}{\res}(\rho) - \wavcoef{s}{\res}(\rho)}^2_{\SO}\right\},
	\label{eq:mse}
\end{align}
and present the result in the following theorem.
\begin{theorem}
	Let the source signal $s(\unit{x})$ be a realization of an anisotropic process which is contaminated by a realization of an additive, zero-mean and anisotropic noise process, $z(\unit{x})$, to obtain the observation $f(\unit{x}) = s(\unit{x}) + z(\unit{x})$. The source and noise signals are uncorrelated, i.e., $\E\{(s)_{\ell}^m \overline{(z)_{p}^q}\} = 0, \, \forall \, \ell, p, |m| \le \ell, |q| \le p$, with known spectral covariance matrices, defined as $C^s_{\ell m,p q} = \E\{(s)_{\ell}^m \overline{(s)_{p}^q}\}$ and $C^z_{\ell m,p q} = \E\{(z)_{\ell}^m \overline{(z)_{p}^q}\}$. Defining $\const$ as in \eqref{eq:WignerD_ortho}, spectral coefficients of the joint $\SO$-scale domain filter in \eqref{eq:Xi}, which minimize the joint $\SO$-scale domain mean square error in \eqref{eq:mse}, can be obtained by inverting the following linear system
	\begin{align}
		\B{A}^{\mathrm{T}}(\ell) \, \B{\Xi}(\res,\ell,m) = \B{b}(\ell,m),
		\label{eq:Xi_matrix}
	\end{align}
	for $|m| \le \ell, \, 0 \le \ell \le L-1, \, \res_1 \le \res \le \res_2$, where elements of the column vector $\B{\Xi}(\res,\ell,m)$ are given $\Xi_k = \left(\Xi(\cdot;\res)\right)^{\ell}_{m,k}, |k| \le \ell$, and elements of the matrix $\B{A}$ and column vector $\B{b}$ are given by
	\begin{align}
		A_{k,k'} = \const \sum_{k=-\ell}^{\ell} \left(C^s_{\ell k,\ell k'} + C^z_{\ell k,\ell k'} \right), \, |k|, |k'| \le \ell,
		\label{eq:A}
	\end{align}
	\begin{align}
		b_{k'} = C^s_{\ell m,\ell k'}, \, |k'| \le \ell.
		\label{eq:b}	
	\end{align}
	\label{th:theorem1}	
\end{theorem}
\vspace{-10mm}
\begin{proof}
	Using \eqref{eq:wavelet_coeff} and \eqref{eq:filtered_wavelet_coeff}, along with the orthogonality of Wigner-$D$ functions on the $\SO$ rotation group, the mean square error in \eqref{eq:mse} can be written as
	\begin{align}
		&\mse = \sum_{\res=\res_1}^{\res_2} \sum_{\ell,m,m'}^{L-1} \const \times \nonumber \\
		& \,\, \E \Bigg\{ \left(\const \sum_{k=-\ell}^{\ell} (f)_{\ell}^k \overline{(\wav{\res})_{\ell}^{m'}} \left(\Xi(\cdot;\res)\right)^{\ell}_{m,k} - (s)_{\ell}^m \overline{(\wav{\res})_{\ell}^{m'}}\right) \times \nonumber \\
		& \quad \overline{\left(\const \sum_{k'=-\ell}^{\ell} (f)_{\ell}^{k'} \overline{(\wav{\res})_{\ell}^{m'}} \left(\Xi(\cdot;\res)\right)^{\ell}_{m,k'} - (s)_{\ell}^m \overline{(\wav{\res})_{\ell}^{m'}}\right)} \Bigg\}.
		\nonumber
	\end{align}
	Differentiating $\mse$ with respect to $\overline{\left(\Xi(\cdot;\res)\right)^{\ell}_{m,k'}}$ and setting the result equal to zero, we obtain a linear system which, using \eqref{eq:A} and \eqref{eq:b}, can be cast in the matrix form in \eqref{eq:Xi_matrix}.
\end{proof}

Having found the spectral representation of the joint $\SO$-scale domain filter, signal estimate $\tilde{s}(\unit{x})$ is obtained from the wavelet coefficients in \eqref{eq:filtered_wavelet_coeff} using \eqref{eq:f_recons_wavelet}.

\begin{remark}
	For wavelet functions which are azimuthally symmetric, i.e., $\left(\wav{\res}\right)_{\ell}^m = \left(\wav{\res}\right)_{\ell}^{0} \delta_{m,0}$, wavelet coefficients are defined on the sphere as~\cite{Leistedt:s2let_axisym}
	\begin{align*}
		\wavcoef{f}{\res}\!(\unit{x}) \!=\! \innerpS{f}{\D_{\unit{x}}\wav{\res}} \!\!= \!\! \sum_{\ell,m}^{\infty} \sqrt{\frac{\const}{2\pi}} (f)_{\ell}^m \overline{(\wav{\res})_{\ell}^0} \,\, Y_{\ell}^m(\unit{x}).
	\end{align*}
	Hence, instead of the joint $\SO$-scale domain filter, we design a joint spatial-scale domain filter as
	\begin{align}
			\Xi(\unit{x};\res) = \sum_{\ell,m}^{L_{\Xi_{\res}}-1} \left(\Xi(\cdot;\res)\right)_{\ell}^m Y_{\ell}^m(\unit{x}),
		\label{eq:Xi_axisym}
	\end{align}
	whose action on the noise-contaminated observation $f$ is given by the following $\untsph$ convolution~\cite{Khalid:2013spie}
	\begin{align}
		\wavcoef{\tilde{s}}{\res}\!\!(\unit{x}) \! = \! \sum_{\ell,m}^{L-1} \! \left(\! \sqrt{\frac{\const}{2\pi}} (f)_{\ell}^m \overline{(\wav{\res})_{\ell}^0}\right) \! \left(\Xi(\cdot;\res)\right)_{\ell}^m  \! Y_{\ell}^m(\unit{x}),
		\label{eq:filtered_wavelet_coeff_axisymm}
	\end{align}
	which necessitates the minimization of the following joint spatial-scale domain mean square error for finding the spectral coefficients of the filter function in \eqref{eq:Xi_axisym}
	\begin{align}
		\mse &= \E\left\{\sum_{\res = \res_1}^{\res_2} \norm{\wavcoef{\tilde{s}}{\res}(\unit{x}) - \wavcoef{s}{\res}(\unit{x})}^2_{\untsph}\right\}.
		\label{eq:mse_axisymm}
	\end{align}
	From the relation between Wigner-$D$ functions and spherical harmonics in \eqref{eq:Wigner_SphHarm_relation}, it can be seen that \eqref{eq:filtered_wavelet_coeff_axisymm} can be obtained from \eqref{eq:filtered_wavelet_coeff} by setting $m' = 0$ and $\left(\Xi(\cdot;\res)\right)^{\ell}_{m,k} = \left(1/\const\right) \left(\Xi(\cdot;\res)\right)_{\ell}^m \delta_{m,k}$. Hence, by setting $k = k' = m$ in \eqref{eq:A} and \eqref{eq:b}, spectral coefficients of the filter in \eqref{eq:Xi_axisym} can be directly obtained from \eqref{eq:Xi_matrix} as
	\begin{align}
		\left(\Xi(\cdot;\res)\right)_{\ell}^m = \frac{C^s_{\ell m,\ell m}}{(C^s_{\ell m,\ell m} + C^z_{\ell m,\ell m})}.
		\label{eq:Xi_axisym_spectral}
	\end{align}
	\label{rem:Remark1}
\end{remark}

\section{Analysis}
\label{sec:analysis}
\begin{figure}[!t]
	\centering
	\vspace{-4.5mm}
	\subfloat[$s(\unit{x})$]
	{		
		\includegraphics[width=0.105\textwidth]{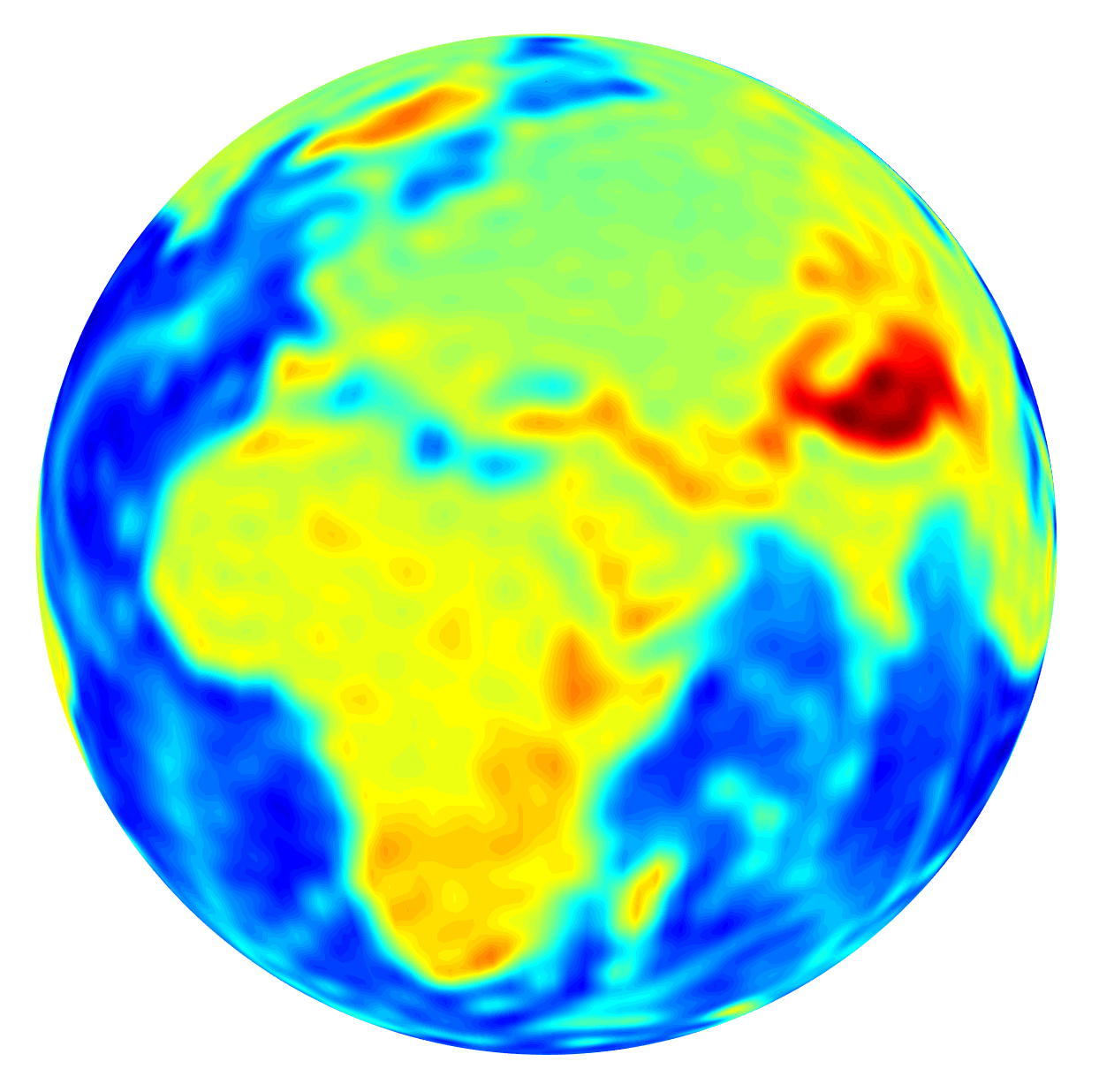}
	}\hfil
	\subfloat[$z(\unit{x})$]
	{		
		\includegraphics[width=0.105\textwidth]{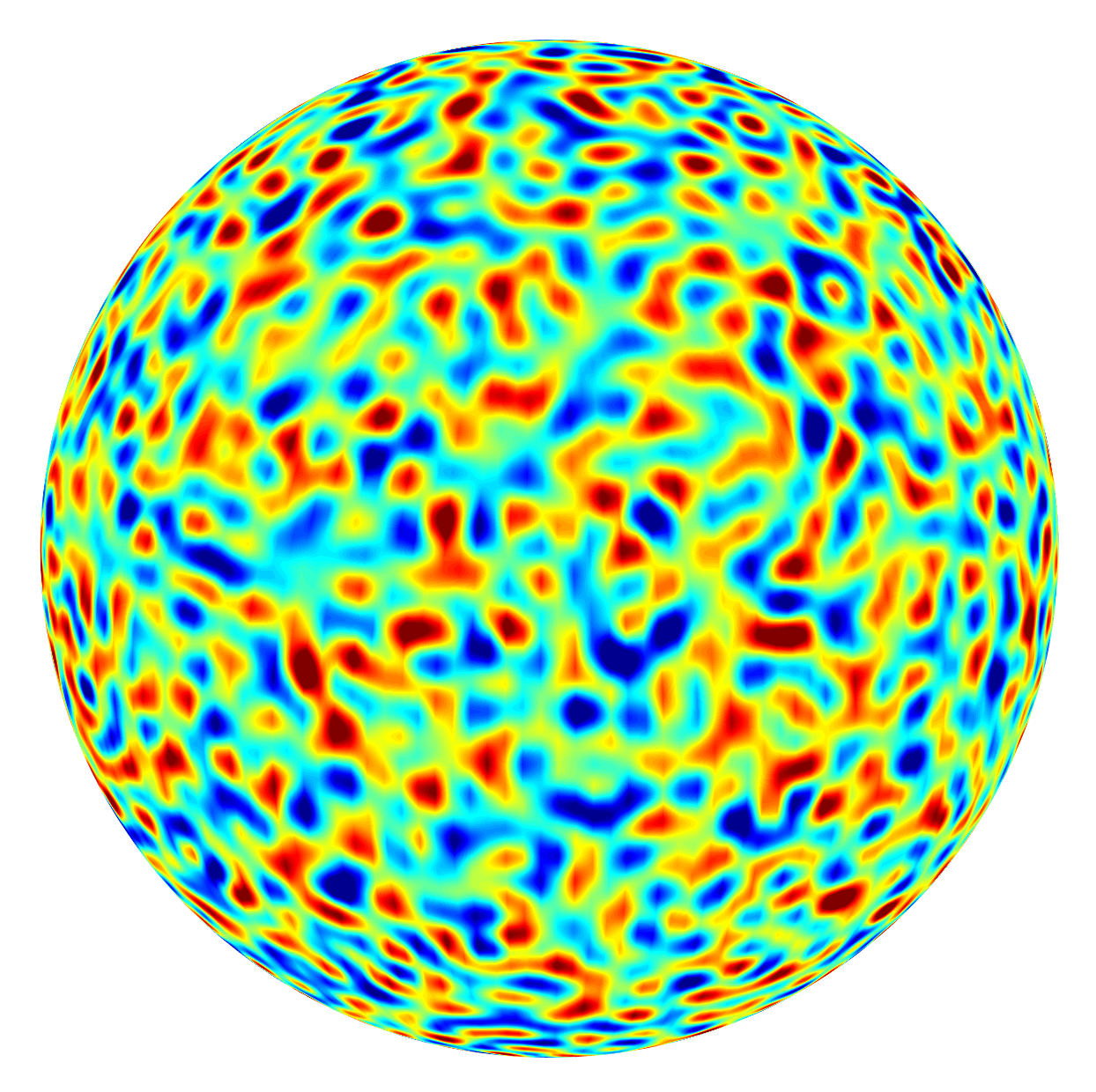}
	}\hfil
	\subfloat[$f(\unit{x})$]
	{
		\includegraphics[width=0.105\textwidth]{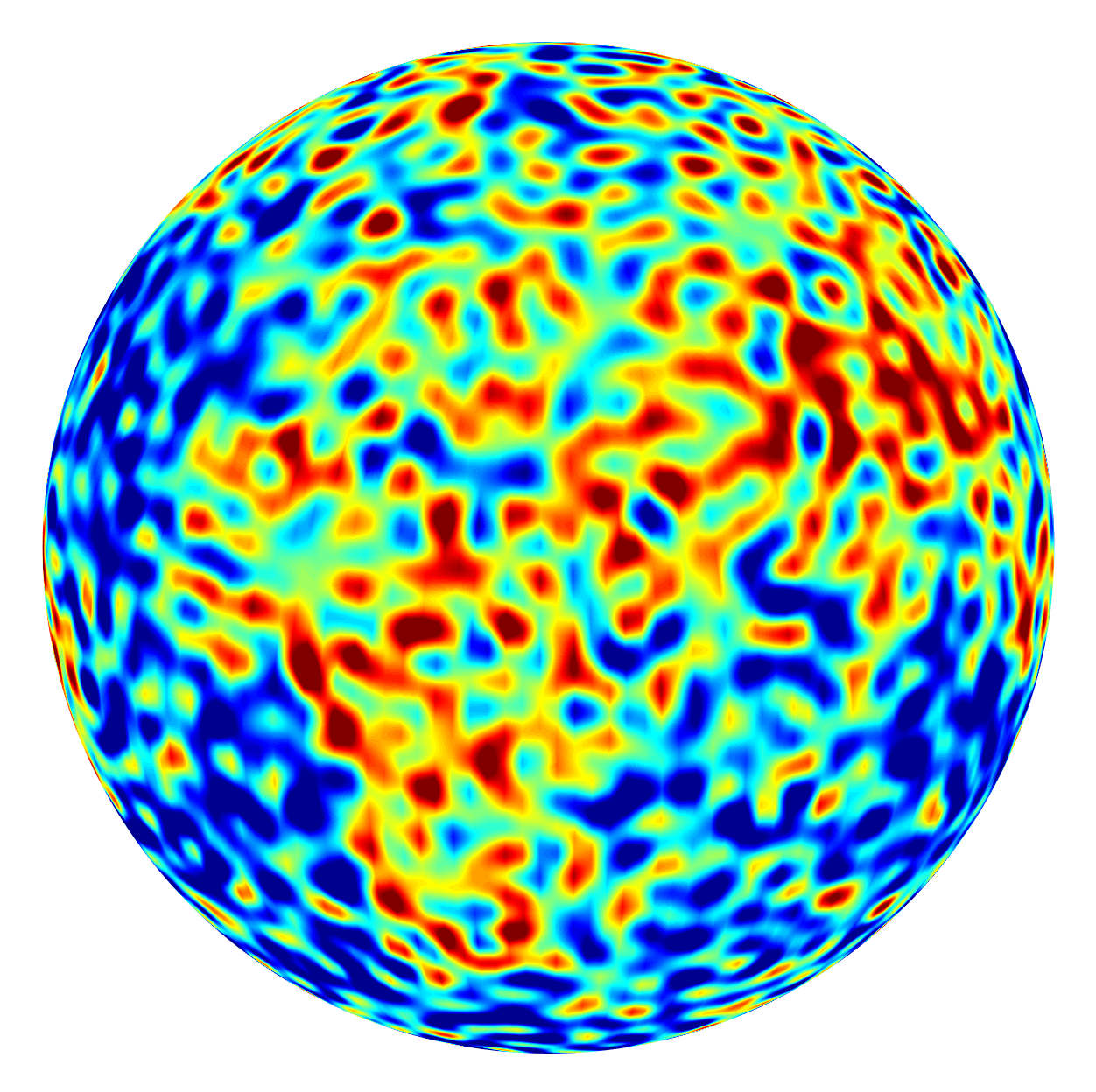}
	}\hfil
	\subfloat[$\tilde{s}(\unit{x})$]
	{
		\includegraphics[width=0.105\textwidth]{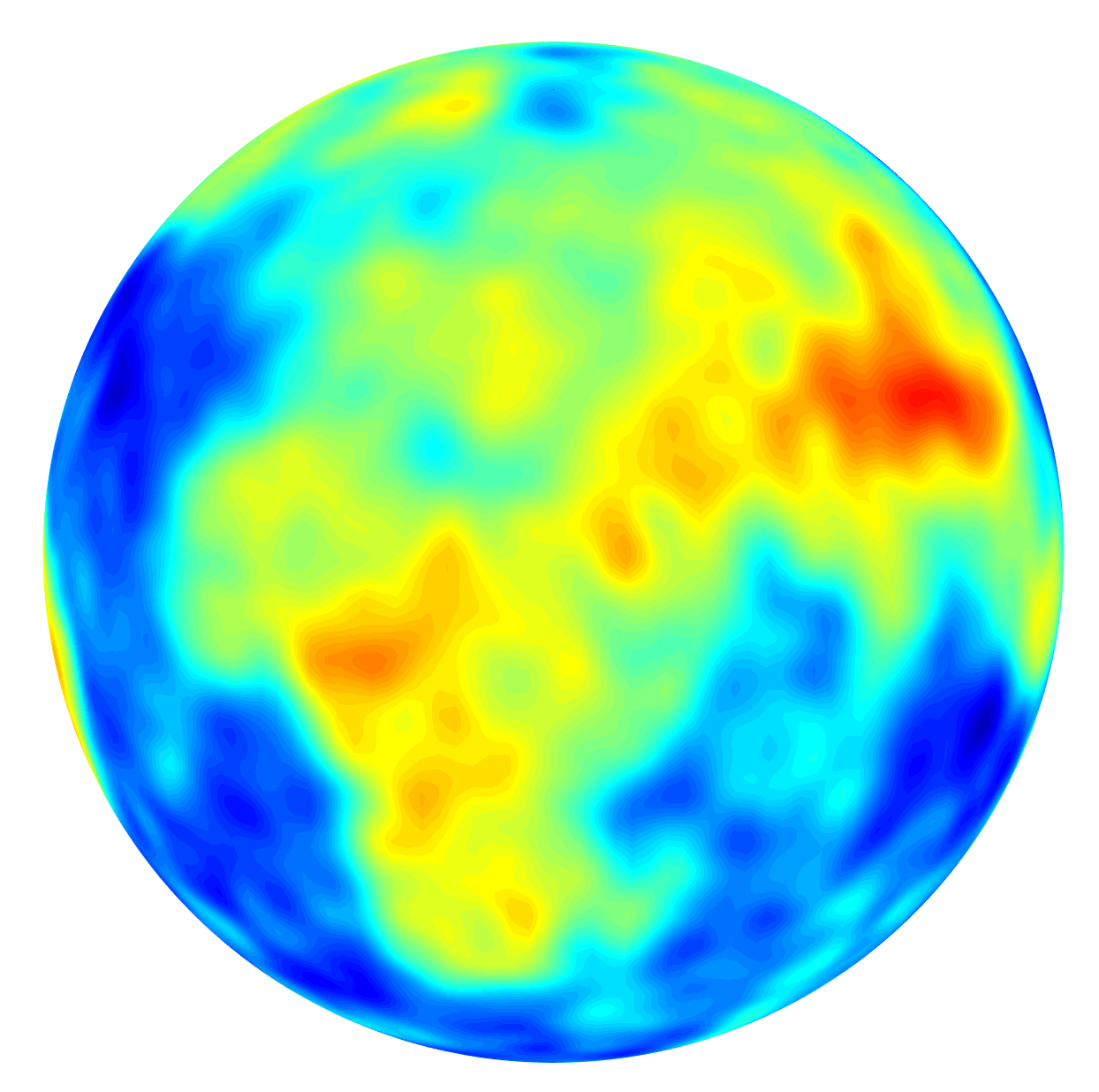}
	}
	
	\vspace{-2mm}
	\subfloat
	{
		\includegraphics[width=0.48\textwidth]{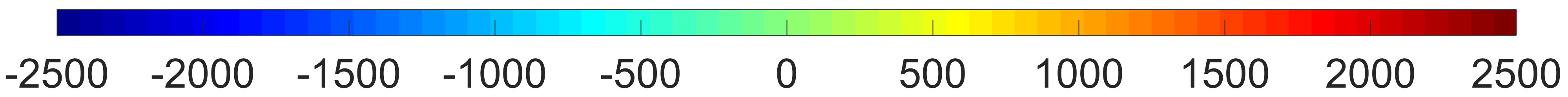}
	}\hfil
	\vspace{-3mm}
	\caption{Multiscale optimal filtering of the bandlimited Earth topography map $s(\unit{x})$, which is contaminated with zero-mean, uncorrelated and white Gaussian noise $z(\unit{x})$ at $\snr{f} = -0.057$ dBs. Output SNR obtained from the source signal estimate $\tilde{s}(\unit{x})$ is $9.68$ dBs, resulting in SNR gain of $9.7$ dBs.}
	\label{fig:denoising}
\end{figure}

Performance of the multiscale optimal filter using scale-discretized wavelet transform is gauged by the signal-to-noise ratio, which for any signal $d \in \lsph$ is defined as
\begin{align}
	\snr{d} = 20 \log \frac{\norm{s(\unit{x})}_{\untsph}}{\norm{d(\unit{x}) - s(\unit{x})}_{\untsph}}.
	\label{eq:snr}
\end{align}
From this definition, input and output SNRs are given by $\snr{f}$ and $\snr{\tilde{s}}$ respectively. We demonstrate the utility of the multiscale optimal filter by using the Earth topography map\footnote{http://geoweb.princeton.edu/people/simons/software.html}, bandlimited to degree $L = 64$, as the source signal $s(\unit{x})$, and contaminating it with realizations of zero-mean, uncorrelated and white Gaussian noise process at different values of $\snr{f}$. Spectral covariance matrix of the source signal is computed as $C^s_{\ell m,p q} = (s)_{\ell}^m \overline{(s)_p^q}$. Spectral covariance matrix of the Gaussian noise process is generated as $\B{C}^z = \sigma^2 \B{I}$, where $\B{I}$ is identity matrix and $\sigma^2$ is the normalized noise energy, i.e., 
\begin{align}
	\sigma^2 = \frac{1}{L^2} 10^{-\snr{f}/10} \, \sum_{\ell,m}^{L-1} \left|(s)_{\ell}^m\right|^2.
	\label{eq:noise_process_energy}
\end{align}
We use azimuthally symmetric wavelet and scaling functions for filtering and estimation of the source signal using the scale-discretized wavelet transform. At $L = 64$, the maximum wavelet scale $\res_2 = 6$~\cite{McEwen:2018}, and we choose the minimum wavelet scale $\res_1 = 0$. Dilation parameter for the harmonic tiling functions is set to $2$~\cite{McEwen:2018}. \figref{fig:denoising} shows an illustration of the optimal filtering framework in which the bandlimited Earth topography map is contaminated with zero-mean, uncorrelated and white Gaussian noise at $\snr{f} = -0.057$ dBs. Output SNR is measured to be $9.68$ dBs, resulting in a significant SNR gain of $9.7$ dBs.
\begin{figure}[!t]
	\centering
	\includegraphics[width=0.49\textwidth]{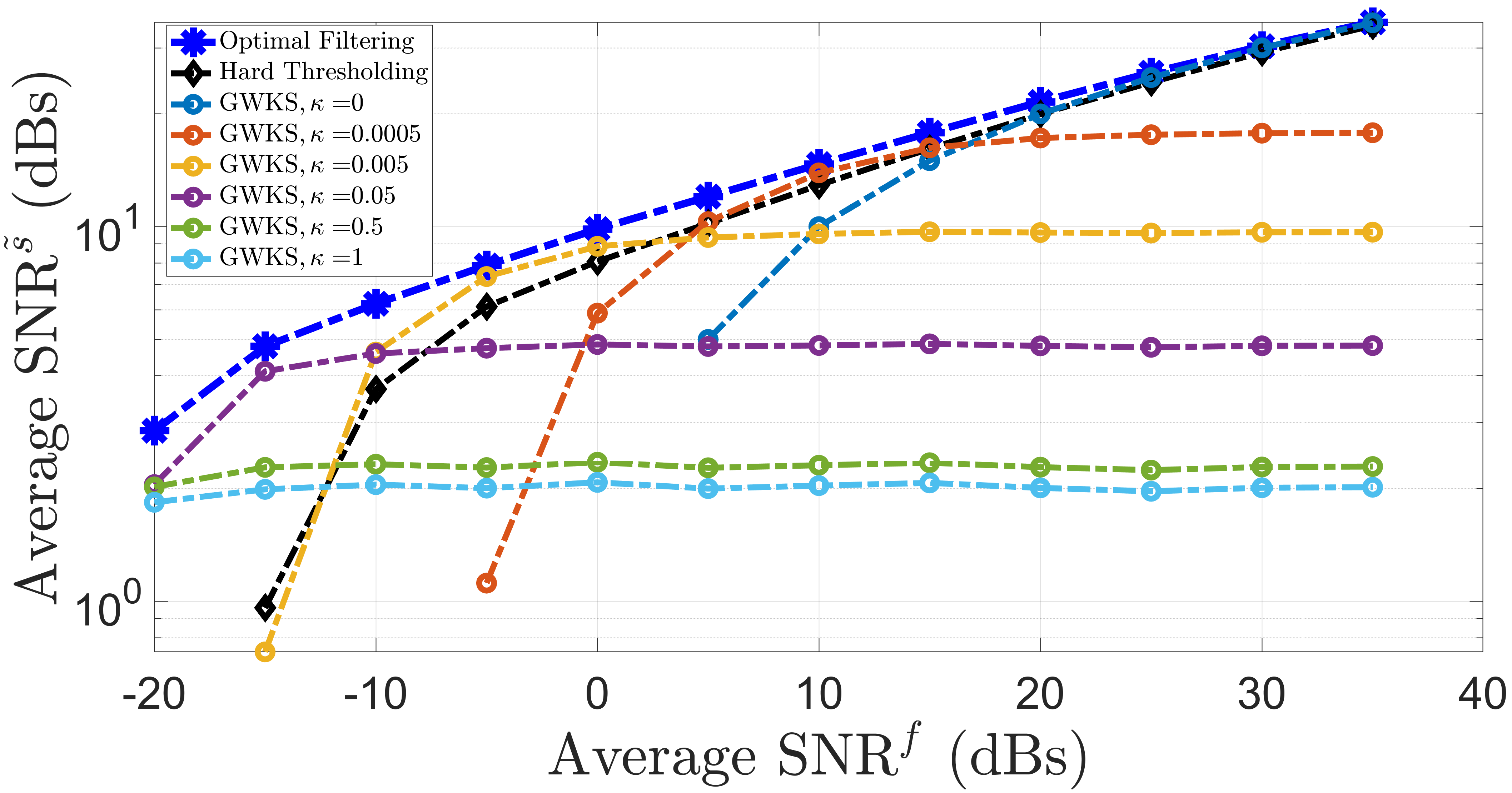}
	\caption{Output SNR plotted against input SNR, averaged over $10$ realizations of zero-mean, uncorrelated and white Gaussian noise process, for the proposed filtering framework~(blue curve), hard thresholding method~(black curve), and weighted-SPHARM based GWKS at different values of $\kappa$.}
	\vspace{-3mm}
	\label{fig:SNR_trend}
\end{figure}

We compare the performance of the proposed filtering framework with the hard thresholding method for signal denoising~\cite{Leistedt:s2let_axisym}, by setting the threshold equal to $3\sigma_{\res}$, where $\sigma^2_{\res} \dfn \E\left\{\left|\wavcoef{z}{\res}(\unit{x})\right|^2\right\}$ is the noise variance in the wavelet domain at scale $\res$ and is given by~\cite{Leistedt:s2let_axisym}
\begin{align}
	\sigma^2_j = \sigma^2 \sum_{\ell=0}^{L-1}\left|\left(\wav{\res}\right)_{\ell}^0\right|^2, \quad \sigma^2 = \frac{\mathrm{trace(\B{C}^z)}}{L^2}.
	\label{eq:Aniso_noise_variance}
\end{align}
We also compare the proposed multiscale optimal filter with the weighted-SPHARM framework presented in~\cite{Chung:2007}, which uses Gauss-Weierstrass kernel smoothing~(GWKS) to obtain the signal estimate by minimizing the following weighted squared error\footnote{We use conjugate of the kernel function in \eqref{eq:kernel_smoothing_criterion} so as to reproduce the results of weighted-SPHARM in~\cite{Chung:2007} using complex spherical harmonics.}
\begin{align}
	\intsph \! \intsph \overline{K_{GW}(\unit{x},\unit{y})} \left|f(\unit{y}) - \tilde{s}(\unit{x})\right|^2 ds(\unit{y}) ds(\unit{x}),
	\label{eq:kernel_smoothing_criterion}
\end{align}
where $K_{GW}(\unit{x},\unit{y})$ is the self-adjoint Gauss-Weierstrass~(GW) kernel given by~\cite{Freeden:1995}
\begin{align}
	K_{GW}(\unit{x},\unit{y}) = \!\! \sum_{\ell,m}^{L-1} e^{-\ell(\ell+1)\kappa} Y_{\ell}^m(\unit{x}) \overline{Y_{\ell}^m(\unit{y})}, \, \kappa \in [0,1].
	\label{eq:GW_kernel}
\end{align}
The signal estimate obtained from the weighted-SPHARM framework is given by~\cite{Chung:2007}
\begin{align}
	\tilde{s}(\unit{x}) = \sum_{\ell,m}^{L-1} e^{-\ell(\ell+1)\kappa} (f)_{\ell}^m Y_{\ell}^m(\unit{x}).
	\label{eq:KS_se_GW_kernel}
\end{align}

We contaminate the Earth topography map, bandlimited to degree $L = 64$, with $10$ realizations of zero-mean, uncorrelated and white Gaussian noise process at different values of $\snr{f}$, and compute $\snr{\tilde{s}}$. \figref{fig:SNR_trend} shows the output SNR versus input SNR, averaged over $10$ realizations\footnote{Average output SNR becomes negative in \figref{fig:SNR_trend}~(and hence, is not shown on the logarithmic scale) at some values of the input SNR for the hard thresholding and weighted-SPHARM methods.}, in which the proposed multiscale filtering method can be seen to perform better than the hard thresholding method, particularly in the low SNR regime. Since there is no systematic way of choosing the GW kernel parameter $\kappa$ for weighted-SPHARM, we estimate the signal at various values of $\kappa$ in the interval $[0,1]$, and show in \figref{fig:SNR_trend} that the proposed multiscale filter outperforms the weighted-SPHARM based GWKS at all values of $\kappa$.

\section{Conclusion}
\label{sec:conclusion}
We have formulated an optimal filter in the wavelet domain, using the scale-discretized wavelet transform, for estimation of signals contaminated by realizations of an additive, zero-mean, uncorrelated and anisotropic noise process on the sphere. The designed filter is optimal in the sense that the filtered representation in the wavelet domain is the minimum mean square error estimate of the wavelet representation of the noise-free signal. We have illustrated the utility of the proposed filtering framework on a bandlimited Earth topography map in the presence of additive, zero-mean, uncorrelated and white Gaussian noise, and have compared the proposed multiscale optimal filtering framework with the hard thresholding and weighted-SPHARM based methods for signal estimation, showing the proposed filter to be superior in performance at different noise levels.

\bibliography{IEEEabrv,sht_bib}

\end{document}